\documentclass[a4paper,10pt]{article}
\usepackage[utf8]{inputenc}

\usepackage{mypackages}
\usepackage{mymacros}

\title{Stability Property for the Call-by-Value $\lambda$-calculus\\ through Taylor Expansion}
\author{Davide Barbarossa \\
Departement of Computer Science, University of Bath}
\date{}

\begin{document}

\maketitle

\begin{abstract}
We prove the \emph{Stability Property} for the \emph{call-by-value} $\lam$-calculus (CbV in the following) (Theorem \ref{cbv-th:TeStability}).
This result states necessary conditions under which the contexts of the CbV $\lam$-calculus 
commute with intersections of approximants.
This is an important non-trivial result, which implies the sequentiality of the calculus (Corollary \ref{cbv-cor:Por}).
We prove it via the tool of resource approximation \cite{DBLP:journals/tcs/EhrhardR08}, whose power has been shown, e.g.\ in \cite{DBLP:journals/pacmpl/BarbarossaM20,DBLP:conf/lics/Barbarossa22,DBLP:journals/lmcs/CerdaA23}.
This technique is usually conceived for the ordinary $\lam$-calculus\footnote{Sometimes, in contrapposition to the CbV $\lambda$-calculus, this is addressed as ``Call by Name''. Strictly speaking however, the ordinary $\lambda$-calculus does \emph{not} follow a supposed ``CbN evaluation'' since there is no restriction on redexes' firing.
}, but it can be easily defined for the CbV setting~\cite{DBLP:journals/corr/abs-1809-02659}.
Our proof is the adaptation of the ordinary one given in \cite{DBLP:journals/pacmpl/BarbarossaM20}, with some minimal technical modification due to the fact that in the CbV setting one linearises terms in a slightly different way than usual (cfr.\ $!(A\multimap B)$ vs $!A\multimap B$).
The content of this contribution is taken from the PhD thesis \cite{DBLP:phd/hal/Barbarossa21}.

\end{abstract}

\section{CbV and its resource approximation in a nutshell}\label{cbv-sec:nutshell}
 
\begin{Definition}[CbV $\lam$-calculus\index{Calculus!CbV $\lam$-!}]
 The set $\Lam_{\cbv}$ of the CbV $\lam$-terms is the same as for the ordinary $\lam$-calculus. The set $\mathrm{Val}$ of \emph{values} contains exactly variables and abstractions.
 $k$-Contexts ($k$ different holes occurring any number of times) are defined as for ordinary $\lam$-calculus.
 The reduction $\vredto\,\subseteq\Lam_{\cbv}\times\Lam_{\cbv}$ of CbV $\lam$-calculus is defined in \cite[Definition 1.5]{DBLP:journals/corr/abs-1809-02659} and it is confluent (\cite[Proposition 1.18]{DBLP:journals/corr/abs-1809-02659}).
\end{Definition}

\begin{Definition}[Resource CbV $\lam$-calculus\index{Resource!-CbV!-$\lam$-calculus}]{\cite[Def.\ 3.1]{DBLP:journals/corr/abs-1809-02659}}\label{cbv-def:cbv-lam}
 The set $\Lam_{\cbv}^{\mathrm{r}}$ of the resource-CbV $\lam$-terms is the set
 $
  \Lam_{\cbv}^{\mathrm{r}}:=\mathrm{Val}^{\mathrm{r}}\,\cup\,\mathrm{Simp}^{\mathrm{r}}
 $
 of \emph{CbV-terms}, where the sets $\mathrm{Val}^{\mathrm{r}}$ of \emph{resource values} and $\mathrm{Simp}^{\mathrm{r}}$ of \emph{resource simple terms} are defined by mutual induction (plus $\alpha$-equivalence) by:
 \[\begin{array}{llllllllllll}
   \mathrm{Val}^{\mathrm{r}}: & v ::= & x \mid \lam x.s
   & & & & & & & 
   \mathrm{Simp}^{\mathrm{r}}: & s ::= & s_1s_2 \mid [v_1,\dots,v_n]
  \end{array}\]
 The reduction $\rredto\subseteq \Lam_{\cbv}^{\mathrm{r}}\!\!\times\!\!\mathscr{P}(\Lam_{\cbv}^{\mathrm{r}})$ of the\! resource\! CbV\! calculus is defined in\! \cite[Definition 3.3, 3.4]{DBLP:journals/corr/abs-1809-02659}.
\end{Definition}


\begin{Proposition}{\cite[Proposition 3.6]{DBLP:journals/corr/abs-1809-02659}}
 The reduction $\rredto$ is confluent and strongly normalising.
\end{Proposition}

\begin{Definition}[Qualitative CbV-Taylor expansion]{\cite[Def.\ 3.9]{DBLP:journals/corr/abs-1809-02659}}
 The \emph{(qualitative) CbV-Taylor expansion}\index{Taylor!qualitative- expansion!cbv $\lam$-calculus-} is the following map $\mathcal{T}:\Lam_{\cbv}\to \mathcal{P}(\mathrm{Simp}^{\mathrm{r}})$:
 \[
   \Te{x} := \set{\,[x,\overset{(n)}{\dots},x]\mid n\in\N} \qquad\qquad\qquad
   \Te{\lam x.M} := \set{\,[\lam x.s_1,\dots,\lam x.s_n] \mid n\in\N\textit{ and }s_i\in\Te{M}}
   \]\[
   \Te{M_1M_2} := \set{s_1s_2 \mid s_i\in\Te{M_i}}.
 \]
 One defines as usual the set $\NFT{M}:=\bigcup_{s\in\Te{M}} \nf{M} \subseteq\nf{\mathrm{Simp}^{\mathrm{r}}}$. 
The inclusion $\NFT M\subseteq \NFT N$ defines as usual a partial preorder $M\leq N$, as well as its symmetric closure equivalence $\NFTeq$.
\end{Definition}

\begin{Remark}
One has $\nf{MN}=\nf{\nf{M}\nf{N}}$ and $\nf{\lam x.M}=\lam x.\nf{M}$, whenever the written normal forms exist.
Moreover, $\NFT{\lam x.M}:=\set{\,[\lam x.s_1,\dots,\lam x.s_n] \mid n\in\N \textit{ and } s_i\in\NFT{M}}$ and $\NFT{MN}:=\bigcup_{s_i\in\NFT{M}} \nf{s_1 s_2}$.
\end{Remark}

The following result 
appears in \cite[Lemma 4.6]{DBLP:journals/corr/abs-1809-02659}. \cite{DBLP:phd/hal/Barbarossa21} gives a more concise inductive proof.

\begin{theorem}[Monotonicity Property\index{Property!Monotonicity- for $\leq$!-in cbv}]\label{cbv-th: Monotonicity}
Any $n$-context $C$ is monotone w.r.t. $\leq$.
\end{theorem}

Remark that Theorem \ref{cbv-th: Monotonicity} implies that contexts are well-defined functions on the quotient $\Lam_{\mathrm{cbv}}/_{=_{\mathcal{T}}}$.

We can simulate $\vredto$ via $\twoheadrightarrow_{\mathrm{r}}$.
As for the ordinary $\lam$-calculus, this is one of the fundamental features of a notion of approximation.

\begin{Proposition}[Simulation Property]{\cite[Lemma 4.4]{DBLP:journals/corr/abs-1809-02659}}\label{cbv-prop:LiftAssSimulation}
If $M\vredto N$ then:
\begin{enumerate}\label{cbv-supposition}
	\item for all $s\in\Te{M}$ there exists $\Sum{T}\subseteq\Te{N}$ s.t.\ $s\msto[\mathrm{r}]\Sum{T}$
	\item for all $s'\in\Te{N}$ s.t.\footnote{The condition $s'\not\to_{0} 0$ refers to a particular reduction $\to_0$, which we did not specify, see \cite{DBLP:journals/corr/abs-1809-02659}.} $s'\not\to_{0} 0$, there exists $s\in\Te{M}$ s.t.\ $s\msto[\mathrm{r}]s'+\Sum{T}$ for some sum $\Sum{T}$.
\end{enumerate}
\end{Proposition}

A CbV $\lam$-theory is a congruence containing the reflexive symmetric and transitive closure $=_{\mathrm{v}}$ of $\vredto$.

\begin{Corollary}[Taylor normal form $\lam$-theory]\label{cbv-cor:lamTh}
The equivalence $\NFTeq$ is a CbV $\lam$-theory.
\end{Corollary}

The proofs of the following results are either trivial or appear in \cite[Lemma 4.9]{DBLP:journals/corr/abs-1809-02659}.

\begin{Remark}\label{cbv-rem:TV=!Val}
 If $V\in \mathrm{Val}$ then $\Te{V}\subseteq \fmsets{\mathrm{Val}^{\mathrm{r}}}$.
\end{Remark}

\begin{Remark}\label{cbv-rem:normalinTe}
 Let $t\in\Lam^{\mathrm{r}}_\mathrm{cbv}$ normal and belonging to $\Te{M}$. If $M\vredto N$ then $t\in\Te{N}$.
\end{Remark}

\begin{Proposition}\label{cbv-prop:unodavantiedietrotuttiquanti}
If $t\in\NFT{M}$, there exists $N\in\Lam_{\mathrm{cbv}}$ s.t.\ $M\msto[\mathrm{v}] N$ and $t\in\Te{N}$.
\end{Proposition}

\begin{Proposition}[Partition Property]\label{cbv-prop:E-R-Lemma}
For all $t,s\in\Te{M}$ s.t.\ $t\neq s$, we have $\nf{t}\cap\nf{s}=\emptyset$.
\end{Proposition}

As for the ordinary $\lam$-calculus, this last property is the key non-trivial ingredient of the proof of the Stability Property. It says that $\NFT M$ is partitioned by the family $\set{\nf{t} \mid t\in\Te{M} \textit{ and }\nf{t}\neq \emptyset}$.

\section{Rigid resource terms}\label{cbv-sec:rigids}

In this section, as for the usual $\lambda$-calculus, we consider ``rigid'' terms/contexts, 
in which we fix an enumeration of the resources appearing in the bags (hence the permutations in the Definitions \ref{cbv-def:Rigids} and \ref{cbv-def:rigidtores}).
This allows us to obtain Lemmas \ref{cbv-lm:NeedInStability} and \ref{cbv-lem:Te_of_contexts}.

\begin{Definition}[CbV resource-contexts]
 The set $\mathrm{Cxt}_{k}^{\mathrm{r}}$ of \emph{CbV resource-$k$-contexts}\index{Contexts!cbv-resource-} is defined as:
  $
  \mathrm{Cxt}_{k}^{\mathrm{r}}:=\mathrm{Val}_k^{\mathrm{r}}\,\cup\,\mathrm{Simp}_k^{\mathrm{r}}
  $,
  where $\mathrm{Val}_k^{\mathrm{r}}$ and $\mathrm{Simp}_k^{\mathrm{r}}$ are defined by mutual induction (\emph{without} $\alpha$-equivalence):
 \[\begin{array}{lllllllllll}
   \mathrm{Val}_k^{\mathrm{r}}: & c^v ::= & \Box_1 \mid \cdots \mid \Box_k \mid x \mid \lam x.c^s
& & & & & & 
\mathrm{Simp}_k^{\mathrm{r}}: & c^s ::= & c^s_1c^s_2 \mid [c^v_1,\dots,c^v_n]
  \end{array}\]
\end{Definition}

We extend the Taylor expansion on each $\mathrm{Cxt}_{k}$ by:
$
 \Te{\Box_i}:=\set{\,[\Box_i,\overset{(n)}{\dots},\Box_i] \mid n\in\N}\subseteq \mathrm{Simp}_k^{\mathrm{r}}.
$

\begin{Definition}[Rigid CbV $\lam$-terms]\label{cbv-def:Rigids}
\begin{enumerate}
\item 
A \emph{rigid $k$-context}\index{Rigid!CbV $\lam$-term} is built as a resource $k$-context but taking lists\footnote{We use $\langle \cdot,\dots,\cdot\rangle$ to denote lists.} of rigid $k$-contexts instead of bags of resource $k$-contexts.
In particular, a rigid term is a rigid context with no occurrences of the holes, taken modulo $\alpha$-equivalence.
As for CbV-terms, rigid contexts are divided into rigid value-contexts and rigid simple-contexts (and this distincion coincides with that of terms when a context has no holes).
\item
Let $c$ be a resource $k$-context. 
We define a set $\mathrm{Rigid}(c)$ of \emph{rigid $k$-contexts} associated with $c$, whose elements are called the \emph{rigids of $c$}\index{Rigid!-of a resource CbV $\lam$-term}, by mutual induction on $\mathrm{Val}_k^{\mathrm{r}}$ and $\mathrm{Simp}_k^{\mathrm{r}}$ as follows:
\[
 \mathrm{Rigid}(\Box_i)=\set{\Box_i}
 \qquad\qquad
 \mathrm{Rigid}(x)=\set{x}
 \]\[
 \mathrm{Rigid}(\lam x.c_0)=\set{\lam x.\linc_0\mid \linc_0\in\mathrm{Rigid}(c_0)}\qquad\qquad
 \mathrm{Rigid}(c_0c_1)=\set{\linc_0\linc_1\mid \linc_i\in\mathrm{Rigid}(c_i)}
 \]\[
 \mathrm{Rigid}([c_1,\dots,c_k])=\set{\langle\linc_{\sigma(1)},\dots,\linc_{\sigma(k)}\rangle\mid \sigma \textit{ permutation and } \linc_i\in\mathrm{Rigid}(c_i)}.
 \]
\end{enumerate}
\end{Definition}

The above definition makes sense since one immediately sees that if $c$ is a resource $k$-value/simple-context then any of its rigids $\linc$ is a rigid $k$-value/simple-context.

\begin{Definition}\label{cbv-def:rigidtores}
Let $\linc$ be a rigid of a CbV resource $k$-context $c$ and, for $i=1,\dots,k$, let $\vec{v}^{\,i}:=\langle v_1^i,\dots,v_{\dg{\Box_i}{c}}^i\rangle$ be a list\footnote{If $\dg{\Box_i}{c}=0$ we mean the empty list.} of resource \emph{values} (that is, elements of $\mathrm{Val}^{\mathrm{r}}$).
We define, by mutual induction on $\mathrm{Val}_k^{\mathrm{r}}$ and $\mathrm{Simp}_k^{\mathrm{r}}$, a resource term $\linc\hole{\vec{v}^1,\dots,\vec{v}^k}\in\Lam^{\mathrm{r}}_{\mathrm{cbv}}$ s.t.\ if $c\in\mathrm{Val}_k^{\mathrm{r}}$ (resp. $\in\mathrm{Simp}_k^{\mathrm{r}}$) then $\linc\hole{\vec{v}^1,\dots,\vec{v}^k}\in\mathrm{Val}^{\mathrm{r}}$ (resp. $\in\mathrm{Simp}^{\mathrm{r}}$).
The definition goes as follows:
\begin{enumerate}
 \item If $c=\Box_i$ then $\linc=\Box_i$; we set $\linc\hole{\langle\rangle,\dots,\langle\rangle,\langle v^i_1\rangle,\langle\rangle,\dots,\langle\rangle}:=v^i_1$
 \item If $c=x$ then $\linc=x$; we set $\linc\hole{\langle\rangle,\dots,\langle\rangle}:=x$
 \item If $c=\lam x.c_0$ then $\linc=\lam x.\linc_0$ where $\linc_0$ is a rigid of $c_0$; we set $\linc\hole{\vec{v}^1,\dots,\vec{v}^k}=\lam x.\linc_0\hole{\vec{v}^1,\dots,\vec{v}^k}$
 \item If $c=c_1c_2$, then $\linc=\linc_1\linc_2$ where $\linc_i$ is a rigid of $c_i$, and each list $\vec{v}^{\,i}$ is a concatenation $\vec{v}^{\,i}=:\vec{w}^{\,i1}\vec{w}^{\,i2}$ where the lists $\vec{w}^{\,ij}$ have exactly $\dg{\Box_i}{c_j}$ elements; we set:
 \[
  \linc\hole{\vec{v}^1,\dots,\vec{v}^k}:=\linc_1\hole{\vec{w}^{11},\dots,\vec{w}^{k1}}\linc_2\hole{\vec{w}^{12},\dots,\vec{w}^{k2}}.
 \]
 \item If $c=[c_1,\dots,c_n]$, then $\linc=\langle \linc_{\sigma(1)},\dots,\linc_{\sigma(n)} \rangle$ where $\sigma$ is a permutation and $\linc_i$ is a rigid of $c_i$, and each list $\vec{v}^{\,i}$ is a concatenation $\vec{v}^{\,i}=:\vec{w}^{\,i1}\cdots\vec{w}^{\,in}$ where the lists $\vec{w}^{\,ij}$ have exactly $\dg{\Box_i}{c_{\sigma(j)}}$ elements;
 we set:
 \[
  \linc\hole{\vec{v}^1,\dots,\vec{v}^k}:=[\linc_{\sigma(1)}\hole{\vec{w}^{\,11},\dots,\vec{w}^{\,k1}},\dots,\linc_{\sigma(n)}\hole{\vec{w}^{\,1n},\dots,\vec{w}^{\,kn}}].
 \]
\end{enumerate}
\end{Definition}

\begin{Remark}
If $v\msto[\mathrm{r}]\Sum V$ then
 $
  \linc\hole{\cdots,\langle \cdots,v,\cdots \rangle,\cdots}\in\Lam^{\mathrm{r}}_{\mathrm{cbv}}
  \msto[\mathrm{r}]
  \{
  \linc\hole{\cdots,\langle \cdots,w,\cdots \rangle,\cdots}
\mid w\in\Sum V\}.
 $
\end{Remark}

The following lemmas will be used in the proof of Theorem \ref{cbv-th:TeStability}.
If $\vec{v}$ is a list we denote with $[\vec{v}]$ the multiset associated with $\vec{v}$ (same elements but disordered).

\begin{Lemma}\label{cbv-lm:NeedInStability}
 Let $C$ be a $k$-context and $c_1,c_2\in\Te{C}$ (hence $c_1,c_2$ are resource $k$-contexts).
 Let $\linc_1$, $\linc_2$ be rigids respectivly of $c_1$, $c_2$.
 For $i=1,\dots,k$, let $\vec{v}^{\,i}=\langle v^i_1,\dots,v^i_{\dg{\Box_i}{c_1}} \rangle$ and $\vec{u}^{\,i}=\langle u^i_1,\dots,u^i_{\dg{\Box_i}{c_2}} \rangle$ be lists of resource \emph{values}.
 If $\linc_1\hole{\vec{v}^{\,1},\dots,\vec{v}^{\,k}}=\linc_2\hole{\vec{u}^{\,1},\dots,\vec{u}^{\,k}}$ then $c_1=c_2$ and $[\vec{v}^{\,i}]=[\vec{u}^{\,i}]$ for all $i$.
\end{Lemma}
\begin{proof}
Induction on $C$.
\begin{enumerate}
 \item[]
 Case $C=\Box_i$.
 Then $c_1=[\Box_i,\overset{(n)}{\dots},\Box_i]$, $c_2=[\Box_i,\overset{(m)}{\dots},\Box_i]$, $\vec{v}^{\,i}=\langle v^{i1},\dots,v^{in}\rangle$, $\vec{u}^{\,i}=\langle u^{i1},\dots,u^{im}\rangle$ and $\vec{v}^{\,j}=\langle\rangle=\vec{u}^{\,j}$ for $j\neq i$.
 So $[v^{i1},\dots,v^{in}]=\linc_1\hole{\vec{v}^1,\dots,\vec{v}^k}=\linc_2\hole{\vec{u}^1,\dots,\vec{u}^k}=[u^{i1},\dots,u^{im}]$, thus $n=m$, i.e. $c_1=c_2$.
 \item[]
 Case $C=x$.
 Then $c_1=[x,\overset{(n)}{\dots},x]$, $c_2=[x,\overset{(m)}{\dots},x]$ and $\vec{v}^{\,i}=\langle\rangle=\vec{u}^{\,i}$.
 So $[x,\overset{(n)}{\dots},x]=\linc_1\hole{\vec{v}^1,\dots,\vec{v}^k}=\linc_2\hole{\vec{u}^1,\dots,\vec{u}^k}=[x,\overset{(m)}{\dots},x]$, thus $n=m$, i.e. $c_1=c_2$.
 \item[]
 Case $C=\lam x.C_0$.
 Then, for $i=1,2$, one has $c_i=[\lam x.c_{i1},\dots,\lam x.c_{in_{i}}]$ with $c_{ij}\in\Te{C_0}$ for all $i,j$.
 So $\linc_i=\langle \lam x.\linc_{i\sigma_i{(1)}},\dots,\lam x.\linc_{i\sigma_i{(n_{i})}}\rangle$ where $\sigma_i$ is a permutation on $n_{i}$ elements.
 By Definition \ref{cbv-def:rigidtores} we have that $\linc_1\hole{\vec{v}^1,\dots,\vec{v}^k}$ and $\linc_2\hole{\vec{u}^1,\dots,\vec{u}^k}$ are equal to:
\[
 [\,\lam x.\linc_{i\widetilde{\sigma}_i{(1)}}\hole{\vec{w}^{\,i11},\dots,\vec{w}^{\,ik1}},\dots,\lam x.\linc_{i\widetilde{\sigma}_i{(n_i)}}\hole{\vec{w}^{\,i1{n_i}},\dots,\vec{w}^{\,ik{n_i}}}]
\]
respectively if $i=1$ or $i=2$, where $\widetilde{\sigma}_i$ is some permutation on $n_i$ elements and where the concatenation $\vec{w}^{ij1}\cdots\vec{w}^{ij{n_i}}$ gives $\vec{v}^{j}$ if $i=1$  and gives $\vec{u}^{j}$ if $i=2$.
 From $\linc_1\hole{\vec{v}^1,\dots,\vec{v}^k}=\linc_2\hole{\vec{u}^1,\dots,\vec{u}^k}$ we get that $n_1=n_2=:n$ and that there exists a permutation $\rho$ on $n$ elements which identifies each term of the bag $\linc_1\hole{\vec{v}^1,\dots,\vec{v}^k}$ with the respective one of the bag $\linc_2\hole{\vec{u}^1,\dots,\vec{u}^k}$, that is, for all $j=1,\dots,n$, one has:
 \[\linc_{1j}\hole{\vec{w}^{\,1\,1\,
 {\widetilde{\sigma_1}^{-1}(j)}
 },\dots,\vec{w}^{\,1\,k\,{\widetilde{\sigma_1}^{-1}(j)}}}
 =
 \linc_{2\rho(j)}\hole{\vec{w}^{\,2\,1\,{\widetilde{\sigma_2}^{-1}(\rho(j))}},\dots,\vec{w}^{\,2\,k\,{\widetilde{\sigma_2}^{-1}(\rho(j))}}}.\]
 The inductive hypothesis gives $c_{1j}=c_{2\rho(j)}$ for all $j=1,\dots,n$ and, putting $h:=\widetilde{\sigma_1}^{-1}(j)$, we have $\vec{w}^{\,1\,i\,h
 }=\vec{w}^{\,2\,i\,
 {\widetilde{\sigma_2}^{-1}(\rho({\widetilde{\sigma_1}(h)}))}
 }$ for all $i=1,\dots,k$. Now, the former equalities give $c_1=c_2$, while the latter give $[\vec{v}^{i}]=[\vec{w}^{1i1}\cdots\vec{w}^{1in}]=[\vec{w}^{2i1}\cdots\vec{w}^{2in}]=[\vec{u}^{i}]$.
 \item[]
 Case $C=C'C''$. Analogous and easier than the above case.\qedhere
\end{enumerate}
\end{proof}

\begin{Lemma}\label{cbv-lem:Te_of_contexts}
 Let $C$ be a $k$-context and $V_1,\dots,V_k\in\mathrm{Val}$. Then:
 \[
  \Te{C\hole{V_1,\dots,V_k}}=\set{\linc\hole{\vec{v}^{\,1},\dots,\vec{v}^{\,k}}\mid c\in\Te{C}, \ \linc\mathrm{ \ rigid \ of \ }c,\,[v_1^i,\dots,v_{\dg{\Box_i}{c}}^i]\in\Te{V_i}}.
 \]
\end{Lemma}
\begin{proof}
 Induction on $C$. Let us show only one of the two base cases, namely the one for $C=\Box_i$.
 \[\begin{array}{rcl}
  & & \set{\linc\hole{\vec{v}^{\,1},\dots,\vec{v}^{\,k}}\mid c\in\Te{C}, \ \linc\mathrm{ \ rigid \ of \ }c,\,[v_1^i,\dots,v_{\dg{\Box_i}{c}}^i]\in\Te{V_i}}
  \\
  & = & \set{\langle\Box_i,\overset{(n)}{\dots},\Box_i\rangle\hole{\langle\rangle,\overset{(i-1)}{\dots},\langle\rangle,\langle v^i_1,\dots,v^i_n \rangle,\langle\rangle,\overset{(k-i)}{\dots},\langle\rangle}\mid n\in\N,\,[v_1^i,\dots,v_n^i]\in\Te{V_i}}
  \\
  & = & \set{\,[v_1^i,\dots,v_n^i]\mid \,[v_1^i,\dots,v_n^i]\in\Te{V_i}} = \Te{V_i} =
  \Te{C\hole{V_1,\dots,V_k}}.
 \end{array}
 \]
\end{proof}

\section{Stability}\label{cbv-sec:stability}

Given $\emptyset\neq\cX\subseteq\Lam_{\mathrm{cbv}}$, we can consider $\bigcap_{M\in\cX} \NFT{M} \subseteq \Lam_{\mathrm{cbv}}^{\mathrm{r}}$, which is the $\inf$ (w.r.t.\ $\leq$) of $\cX$ in $\Lam_{\mathrm{cbv}}/\!\NFTeq$.
Note that even if an $M$ such that $M\NFTeq \inf \cX$ exists, it need not be unique in $\Lam_{\mathrm{cbv}}$, but it is unique in $\Lam_{\mathrm{cbv}}/_{=_{\mathcal{T}}}$.
If it exists, we say that $\inf \cX$ \emph{is definable in} $\Lam_{\mathrm{cbv}}/_{=_{\mathcal{T}}}$.
The proof presented below is an adaptation of the proof for the ordinary $\lam$-calculus \cite{DBLP:journals/pacmpl/BarbarossaM20}.
Knowing that the proof technique can be adapted also for the $\lambda\mu$-calculus \cite{DBLP:conf/lics/Barbarossa22}, the fact that one can adapt it for CbV as well, can be seen as yet an extra strength of Taylor resource approximation.
We conclude this section with Corollary \ref{cbv-cor:Por}, which expresses a notion of sequentiality of the calculus.

\begin{theorem}[Stability Property\index{Property!Stability-!for $\Lam_{\mathrm{cbv}}/_{\NFTeq}$}]\label{cbv-th:TeStability} 
Let $C:\Lam_{\mathrm{cbv}}/_{=_{\mathcal{T}}}\times\overset{(n)}{\dots}\times\Lam_{\mathrm{cbv}}/_{=_{\mathcal{T}}}\to\Lam_{\mathrm{cbv}}/_{=_{\mathcal{T}}}$ be an $n$-context and fix non-empty $\cX_1,\dots,\cX_n\subseteq\mathrm{Val}/_{=_{\mathcal{T}}}$, each upper bounded by some value.
If $\inf{\cX_i}$ is definable in $\Lam_{\mathrm{cbv}}/_{=_{\mathcal{T}}}$ by a value for all
$i$, then the $\inf$ of the image of $C$ on $\cX_1\times\overset{(n)}{\dots}\times\cX_n$ is definable in $\Lam_{\mathrm{cbv}}/_{=_{\mathcal{T}}}$ and it is:
\[
C\hole{\inf_{N_1\in\cX_1} N_1,\dots,\inf_{N_n\in\cX_n} N_n} = \inf_{\underset{N_n\in\cX_n}{\underset{\dots}{N_1\in\cX_1}}} C\hole{N_1,\dots,N_n}.
\]
\end{theorem}
\begin{proof}
Wlog we can consider that the $\cX_i$'s are in $\mathrm{Val}$. Since they are upper bound by a value, for $i=1,\dots,n$ there exists $L_i\in\mathrm{Val}$ s.t.\ $\bigcup_{N\in\cX_i}\NFT{N}\subseteq\NFT{L_i}$.
Since the $\inf{\cX_i}$'s are definable in $\Lam_{\mathrm{cbv}}/_{=_{\mathcal{T}}}$ by a value, let $V_1,\dots,V_n\in\mathrm{Val}$ s.t.\ $\NFT{V_i} = \bigcap_{N\in\cX_i}\NFT{N}$.
It suffices now to show that
$
	\NFT{C\hole{V_1,\dots,V_n}} = \bigcap_{N_1\in\cX_1}\dots\bigcap_{N_n\in\cX_n}\NFT{C\hole{N_1,\dots,N_n}}.
$

$(\subseteq)$.
Immediate by \autoref{cbv-th: Monotonicity}.

$(\supseteq)$.
Let $t\in\bigcap\limits_{\vec N\in\vec \cX}\NFT{C\hole{N_{1},\dots,N_{n}}}$ (where $\vec N:=(N_1,\dots,N_n)$ and $\vec{\cX}:=(\cX_1,\dots,\cX_n)$).
For every $\vec N\in\vec\cX$, by Lemma~\ref{cbv-lem:Te_of_contexts} there exists a CbV $n$-resource-context $c_{\vec N}\in\Te{C}$ and, for every $i=1,\dots,n$, a list $\vec v^{\,i}_{\vec N} = \langle v^{i1}_{\vec N},\dots,v^{id_i}_{\vec N} \rangle$ (where $d_i:=\dg{\Box_i}{c_{\vec N}}$) with $[\vec v^{\,i}_{\vec N}]\in\Te{N_{i}}$ and such that $t\in\nf{\linc_{\vec N}\hole{\vec v^{\,1}_{\vec N},\dots,\vec v^{\,n}_{\vec N}}}$, for $\linc_{\vec{N}}$ a rigid of $c_{\vec{N}}$. Confluence allows to factorize the reduction from $\linc_{\vec N}\hole{\vec v^{\,1}_{\vec N},\dots,\vec v^{\,n}_{\vec N}}$ to $t$ as follows:
\[
\linc_{\vec N}\hole{\nf{v^{11}_{\vec N}},\dots,\nf{v^{1d_1}_{\vec N}},\dots,\nf{v^{n1}_{\vec N}},\dots,\nf{v^{nd_n}_{\vec N}}}\\[1ex]
\msto[\mathrm{r}] \nf{\linc_{\vec N}\hole{\vec v^{\,1}_{\vec N},\dots,\vec v^{\,n}_{\vec N}}}
\ni t.
\]
Wlog $d_i\geq 1$ for all $i$ and $\nf{v^{ij}_{\vec N}}\neq\emptyset$ for all $i,j$.
In fact, the holes s.t. $d_i=0$ can be ignored, and if both $d_i\geq 1$ and $\nf{v^{ij}_{\vec N}}\neq\emptyset$ for some $j$, then by the previous line we would have $t\in\nf{\emptyset}$, a contraddiction.

So for all $i=1,\dots,n$ and $j=1,\dots,d_i$, there exists $w^{ij}_{\vec N}\in\nf{v^{ij}_{\vec N}}$ such that:
\begin{equation}\label{cbv-eq:tindoppiavvu}
 \nf{\linc_{\seq N}\hole{\vec w^{\,1}_{\vec N},\dots,\vec w^{\,n}_{\vec N}}} \ni t
\end{equation} 
and being $N_i\in\cX_i$ which is bounded by $L_i$, we have $[\vec{w}^{i}_{\vec N}]\in\nf{[\vec{v}^{i}_{\vec N}]}\subseteq\NFT{N_{i}}\subseteq\NFT{L_i}$.
From the inclusion $[\vec{w}^i_{\vec N}]\in\NFT{L_i}$ we obtain, thanks to Remark \ref{cbv-rem:TV=!Val} because $L_i$ is a value, a simple term $[\vec{u}^{\,i}_{\vec N}]\in\Te{L_i}$ such that:
\begin{equation}\label{cbv-eq:crucialforfinite}
 [\vec{w}^{i}_{\vec N}]\in\nf{[\vec{u}^{i}_{\vec N}]}
\end{equation}
i.e. they have the same number of elements and $\nf{u^{ij}_{\vec N}} \ni w^{ij}_{\vec N}$ for all $i,j,\vec{N}$.
By composing thus a reduction from $\linc_{\seq N}\hole{\vec w^{\,1}_{\vec N},\dots,\vec w^{\,n}_{\vec N}}$ to $t$ with a reduction from $u^{ij}_{\vec N}$ to $w^{ij}_{\vec N}$, we find that $t\in\nf{\linc_{\vec N}\hole{\vec u^{\,1}_{\vec N},\dots,\vec u^{n}_{\vec N}}}$. 
This holds for all $\vec N\in\vec\cX$, i.e.:
\begin{equation}\label{cbv-eq:tinintersubsubsectionu}
 t\in\bigcap_{\vec N\in\vec\cX} \nf{c_{\vec N}\hole{\vec u^{\,1}_{\vec N},\dots,\vec u^{n}_{\vec N}}}.
\end{equation}
Now, Lemma~\ref{cbv-lem:Te_of_contexts} gives $\linc_{\vec N}\hole{\vec u^{\,1}_{\vec N},\dots,\vec u^{n}_{\vec N}}\in\Te{C\hole{L_1,\dots,L_n}}$.
But since the $L_i$'s are independent from $N_1,\dots,N_n$, and thanks to \eqref{cbv-eq:tinintersubsubsectionu}, we can apply Proposition~\ref{cbv-prop:E-R-Lemma}, and obtain that the set  $\set{\linc_{\vec N}\hole{\vec u^{\,1}_{\vec N},\dots,\vec u^{n}_{\vec N}} \st\ \vec N\in\vec \cX}$ is actually a singleton.
Therefore, Lemma \ref{cbv-lm:NeedInStability} tells us that also the terms $\linc_{\vec N}$ and the bags $[\vec u_{\vec N}^{\,i}]$ are independent from $\vec N\in\vec \cX$.
The unique element of the previous sigleton has hence shape $\linc\hole{\vec u^{\,i},\dots,\vec u^{n}}$, with $\linc$ a rigid of a $c\in\Te{C}$, and $[\vec{u}^{\,i}]\in\Te{L_i}$.
Recalling now that $[\vec{w}^{\,i}_{\vec N}] \in \NFT{L_i}$, we can apply Proposition~\ref{cbv-prop:unodavantiedietrotuttiquanti} in order to obtain, for each $i=1,\dots,n$, an $L'_{[\vec{w}^{\,i}_{\vec N}]}\in\Lam_{\mathrm{cbv}}$ such that $L_i\msto[\mathrm{v}]L'_{[\vec{w}^{\,i}_{\vec N}]}$ and $[\vec{w}^{\,i}_{\vec N}] \in\Te{L'_{[\vec{w}^{\,i}_{\vec N}]}}$.
Remark that these $L_{[\vec{w}^{\,i}_{\vec N}]}$'s must in fact be values, since they are reducts of the values $L_i$.
Consider now the set $\set{[\vec{w}^{\,i}_{\vec N}] \mid \vec N\in\vec\cX}$, which can be \emph{a priori} infinite.
Since for $i$ fixed, the set $\set{[\vec{u}^{\,i}_{\vec N}]\mid \vec N\in\vec \cX}$ is a singleton $\set{[\vec{u}^{\,i}]}$, \eqref{cbv-eq:crucialforfinite} entails that $[\vec{w}^{i}_{\vec N}]\in\nf{[\vec{u}^{i}]}$, and our set $\set{[\vec{w}^{\,i}_{\vec N}] \mid \vec N\in\vec\cX}$ must thus in fact be finite.
Therefore we can invoke confluence in order to say that the \emph{finitely many} $L'_{[\vec{w}^{\,i}_{\vec N}]}$'s share a commond reduct, call it $L'_i$, which as the notation shows is now independent from $[\vec{w}^{\,i}_{\vec N}]$ (but still depends on $i$).
Of course $L'_i$ is also a reduct of $L_i$, and it is still a value.
Also, since each $[\vec{w}^{\,i}_{\vec N}]$ belongs to $\Te{L'_{[\vec{w}^{\,i}_{\vec N}]}}$ and is normal, by Remark \ref{cbv-rem:normalinTe} we have $\set{[\vec{w}^{\,i}_{\vec N}] \mid \vec N\in\vec\cX}\subseteq\Te{L'_i}$.
Thus we can apply Lemma~\ref{cbv-lem:Te_of_contexts} and find that, for every $\vec N\in\vec\cX$, we have:
\begin{equation}\label{cbv-eq:TaylorLprimes}
	\linc\hole{\vec w^{\,1}_{\vec N},\dots,\vec w^{n}_{\vec N}}\in\Te{C\hole{L'_1,\dots,L'_n}}.
\end{equation}
But now thanks to \eqref{cbv-eq:tindoppiavvu} (which holds for all $\vec N\in\vec\cX$) and \eqref{cbv-eq:TaylorLprimes}, we can apply again Proposition~\ref{cbv-prop:E-R-Lemma} in order to find that the set $\set{\linc\hole{\vec w^{\,1}_{\vec N},\dots,\vec w^{n}_{\vec N}}\st\ \vec N\in\vec\cX}$ is a singleton.
Again by Lemma \ref{cbv-lm:NeedInStability}, we have that all the bags $[ \vec w^{\,1}_{\vec N}],\dots,[\vec w^{n}_{\vec N} ]$ for $\vec{N}\in\vec{\cX}$, coincide respectively to some bags $[ \vec w^{\,1}],\dots,[\vec w^{n} ]$ which are independent from $\vec N\in\vec\cX$.
So the only element of the previous singleton has shape $\linc\hole{\vec w^{\,1},\dots,\vec w^{n}}$, and by \eqref{cbv-eq:tindoppiavvu} we get:
\begin{equation}\label{cbv-eq:tindoppiavvubis}
 t\in\nf{\linc\hole{\vec w^{\,1},\dots,\vec w^{n}}}.
\end{equation}
Now, for all $i$, remembering what we found already, we have $[\vec{w}^{\,i}]=[\vec{w}^{\,i}_{\vec N}]\in\NFT{N}$ for all $N\in\cX_i$.
That is,
\begin{equation}\label{cbv-eq:endstability}
 [\vec{w}^{\,i}]\in\bigcap_{N\in\cX_i} \NFT{N} =\NFT{V_i}
\end{equation}
where we finally used the hypothesis.
From \eqref{cbv-eq:endstability} and Lemma~\ref{cbv-lem:Te_of_contexts} one can now easily conclude that $t\in\nf{\linc\hole{\vec w^1,\dots,\vec w^n}}\subseteq\NFT{C\hole{V_1,\dots,V_n}}$.
\end{proof}

As usual, one obtains as corollary the non-existence of the following \emph{parallel-or}\index{Parallel-or (all versions)}.
We use the usual encoding of pairs: $(M,N):=\lam z.zMN$.
Remark that a pair is a value.

\begin{Corollary}[No parallel-or]\label{cbv-cor:Por}
There is \emph{no} $\prog{Por}\in\Lam_{\mathrm{cbv}}$ s.t.\ for all $M,N\in\Lam_{\mathrm{cbv}}$,
\[
\left\{
\begin{array}{rlll}
\prog{Por}\,(M,N) & \NFTeq & \prog{True} & \mathrm{if \ }M\not\NFTeq\Om\mathrm{ \ or \ }N\not\NFTeq\Om \\
\prog{Por}\,(M,N) & \NFTeq & \Om & \mathrm{if \ }M\NFTeq N\NFTeq\Om.
\end{array}
\right.
\]
\end{Corollary}
\begin{proof}
Otherwise, for $C:=\prog{Por}\,\Box$, $\cX=\set{(\prog{True},\Om), (\Om,\prog{True})}$ (upper bounded by the value $(\prog{True},\prog{True})$), and the value $V=(\Om,\Om) \NFTeq \inf \set{(\prog{True},\Om), (\Om,\prog{True})}$, Theorem \ref{cbv-th:TeStability} would give the contradiction:
\[
 \prog{True} \NFTeq \inf\set{C\hole{(\prog{True},\Om)}, C\hole{(\Om,\prog{True})}} \NFTeq C\hole{(\Om,\Om)} \NFTeq \Om.\qedhere
\]
\end{proof}

\begin{Remark}
Here $\Om$ is taken as representative of ``operationally meaningless'' term in the calculus. However, one should see what happens with the more appropriate CbV notion studied in \cite{DBLP:journals/corr/abs-2401-12212}.
\end{Remark}

\section{Final comments}

The CbV $\lambda$-calculus that we have used is of the form given in \cite{DBLP:journals/corr/abs-1809-02659}.
However, one could argue that the canonical formulation of CbV should be on the lines of the one given in \cite{DBLP:journals/corr/abs-2401-12212} (explicit substitutions and a distant action reduction). The first work would therefore be to reproduce our proof for that syntax (or at least prove that Stability in either setting is equivalent to Stability in the other).

Moreover, we remark that \autoref{cbv-sec:rigids}, where we have to consider lists instead of multisets, is quite annoying.
This detour also appears, in the exact same form, for the ordinary setting. 
If one would directly define resource approximation with lists (usually called the \emph{rigid/polyadic resource approximation} \cite{DBLP:journals/lmcs/OlimpieriA22,DBLP:journals/pacmpl/MazzaPV18}), this annoying detour would probably disappear (its content would still be there, but in a different shape).
The second work would therefore be to reformulate the whole theory of approximation of CbV (and even the ordinary one!) in a rigid/polyadic way.

Finally, in the recent \cite{DBLP:journals/corr/abs-2401-12212}, a proof of the Genericity Property for CbV is given, in a sense inspired from the one for the ordinary $\lam$-calculus given in \cite{DBLP:journals/pacmpl/BarbarossaM20}. Once a good notion of CbV-B\"ohm trees (or similar) at hand, one should be able to directly adapt the latter proof to CbV and understand the relations between the two proof techniques.
We see the Stability and Genericity Properties of CbV (once agreed on an established form of the calculus) as the first steps of the development of a ``mathematical theory of CbV'', in the same sense that we have for the ordinary one. A third work would be, for instance, to ask if CbV does enjoy the Perpendicular Lines Property and the Continuity Lemma (see \cite{DBLP:journals/pacmpl/BarbarossaM20}).

\bibliographystyle{alpha}
\bibliography{MyBibTeX}

\begin{thebibliography}{KMP20}

\bibitem[AGK24]{DBLP:journals/corr/abs-2401-12212}
Victor Arrial, Giulio Guerrieri, and Delia Kesner.
\newblock Genericity through stratification.
\newblock {\em CoRR}, abs/2401.12212, 2024.

\bibitem[Bar21]{DBLP:phd/hal/Barbarossa21}
Davide Barbarossa.
\newblock {\em Towards a resource based approximation theory of programs. (Vers
  une th{\'{e}}orie de l'approximation des programmes bas{\'{e}}e sur la notion
  de ressources)}.
\newblock PhD thesis, Paris 13 University, Villetaneuse, France, 2021.

\bibitem[Bar22]{DBLP:conf/lics/Barbarossa22}
Davide Barbarossa.
\newblock Resource approximation for the {\(\lambda\)}{\(\mu\)}-calculus.
\newblock In Christel Baier and Dana Fisman, editors, {\em {LICS} '22: 37th
  Annual {ACM/IEEE} Symposium on Logic in Computer Science, Haifa, Israel,
  August 2 - 5, 2022}, pages 27:1--27:12. {ACM}, 2022.

\bibitem[BM20]{DBLP:journals/pacmpl/BarbarossaM20}
Davide Barbarossa and Giulio Manzonetto.
\newblock Taylor subsumes {Scott, Berry, Kahn and Plotkin}.
\newblock {\em Proc. {ACM} Program. Lang.}, 4({POPL}):1:1--1:23, 2020.

\bibitem[CA23]{DBLP:journals/lmcs/CerdaA23}
R{\'{e}}my Cerda and Lionel~Vaux Auclair.
\newblock Finitary simulation of infinitary $\beta$-reduction via taylor
  expansion, and applications.
\newblock {\em Log. Methods Comput. Sci.}, 19(4), 2023.

\bibitem[ER08]{DBLP:journals/tcs/EhrhardR08}
Thomas Ehrhard and Laurent Regnier.
\newblock Uniformity and the {T}aylor expansion of ordinary lambda-terms.
\newblock {\em Theor. Comput. Sci.}, 403(2-3):347--372, 2008.

\bibitem[KMP20]{DBLP:journals/corr/abs-1809-02659}
Axel Kerinec, Giulio Manzonetto, and Michele Pagani.
\newblock Revisiting call-by-value {B\"{o}}hm trees in light of their {T}aylor
  expansion.
\newblock {\em Log. Methods Comput. Sci.}, 16(3), 2020.

\bibitem[MPV18]{DBLP:journals/pacmpl/MazzaPV18}
Damiano Mazza, Luc Pellissier, and Pierre Vial.
\newblock Polyadic approximations, fibrations and intersection types.
\newblock {\em Proc. {ACM} Program. Lang.}, 2({POPL}):6:1--6:28, 2018.

\bibitem[OA22]{DBLP:journals/lmcs/OlimpieriA22}
Federico Olimpieri and Lionel~Vaux Auclair.
\newblock On the taylor expansion of {\(\lambda\)}-terms and the groupoid
  structure of their rigid approximants.
\newblock {\em Log. Methods Comput. Sci.}, 18(1), 2022.

\end{thebibliography}

\end{document}